\documentclass[aps,pra,twocolumn,superscriptaddress,nofootinbib]{revtex4-2}

\usepackage[braket, qm]{qcircuit}
\usepackage{graphicx}
\usepackage{amssymb}
\usepackage{indentfirst}
\usepackage{amsmath}
\usepackage{color,xcolor}
\usepackage{subfigure}
\usepackage{verbatim}
\usepackage[noend]{algpseudocode}
\usepackage{algorithm,algorithmicx}
\usepackage{amsthm}

\usepackage{tikz}

\newtheorem{theorem}{Theorem}

\newtheorem{proposition}{Proposition}
\newtheorem{definition}{Definition}
\newtheorem*{remark}{Remark}

\begin{document}

\title{Classical shadow tomography with mutually unbiased bases}

\author{Yu Wang}
\email[]{ming-jing-happy@163.com}
\affiliation{Beijing Institute of Mathematical Sciences and Applications, Beijing 101408, China}

\author{Wei Cui}
\email[]{cwei@bimsa.cn}

\affiliation{Beijing Institute of Mathematical Sciences and Applications, Beijing 101408, China}
\affiliation{Yau Mathematical Sciences Center, Tsinghua University, Beijing 100084, China}

\begin{abstract}

Classical shadow tomography, harnessing randomized informationally complete (IC) measurements, provides an effective avenue for predicting many properties of unknown quantum states with sample-efficient precision. Projections onto $2^n+1$ mutually unbiased bases (MUBs) are widely recognized as minimal and optimal IC measurements for full-state tomography. We study how to use MUBs circuits as the ensemble in classical shadow tomography.
For the general observables, the variance to predict their expectation value is shown to be exponential to the number of qubits $n$. However, for a special class termed as appropriate MUBs-average (AMA) observables, the variance decreases to $poly(n)$.
Additionally, we find that through biased sampling of MUBs circuits, the variance for non-AMA observables can again be reduced to $poly(n)$ with the MUBs-sparse condition.
The performance and complexity of using the MUBs and Clifford circuits as the ensemble in the classical shadow tomography are compared in the end.

\end{abstract}

\maketitle

\section{Introduction}

In the realm of quantum information science, efficiently extracting information from unknown quantum states is pivotal. This is traditionally achieved through quantum state tomography \cite{leonhardt1995quantum,james2001measurement,paris2004quantum}, performing informationally complete (IC) measurements usually projected on  $\{U_j|k\rangle,k=0,\cdots,d-1;j=1\cdots\}$, obtaining experimental data  $\text{tr}(\rho U_j|k\rangle \langle k| U_j^{\dag})$, then uniquely reconstructing density matrix $\rho$ with different methods. It allows for the prediction of various important functions $f(\rho)$, such as predicting the properties $\text{tr}(O\rho)$ under certain observable $O$, along with metrics like purity and entropy \cite{da2011practical,flammia2011direct,brydges2019probing}. These predictions are central to many-body physics and quantum information theory \cite{preskill2018quantum,amico2008entanglement}. However, as quantum systems scale up, such as in the case of $n$-qubit quantum systems with dimension $d=2^n$, this method becomes impractical and even infeasible due to the enormous memory requirements.

Nevertheless, when computing specific functions $f(\rho)$, we can avoid the need to accurately calculate all elements of the density matrix with exponential measurements. While shadow tomography was initially proposed with polynomial sampling \cite{aaronson2018shadow}, it required exponential-depth quantum circuits applied to copies of all quantum states, presenting challenges for quantum hardware. Subsequently, Huang et al. introduced classical shadow tomography \cite{huang2020predicting}, enabling random measurements on individual quantum states and efficient prediction of various properties with a sampling complexity of ~$\log(M)\|\cdot \|_{\text{norm}}^2$, where $M$ represents the number of observables, and $\|\cdot \|_{\text{norm}}^2$ denotes the norm of the corresponding observables.
This norm is also influenced by the unitary ensemble $\{U_j\}$ we randomly choose.

 The initial procedure applies random unitaries from a specific IC  ensemble to the system and then performs computational projective measurements, which is equivalent to performing randomly $3^n$ Pauli measurements or all Clifford measurements. Pauli measurements are ideal for predicting localized target functions, while Clifford measurements excel in estimating functions with constant Hilbert-Schmidt norms, both offering valuable tools for various quantum tasks. Subsequently, various other ensembles have been explored, including fermionic Gaussian unitaries \cite{zhao2021fermionic}, chaotic Hamiltonian evolutions \cite{hu2022hamiltonian}, locally scrambled unitary ensembles \cite{hu2023classical,akhtar2023scalable,ippoliti2023operator}, and Pauli-invariant unitary ensembles \cite{bu2022classical}. The concept of randomly selecting multiple sets of projective measurements has been theoretically generalized to one POVM \cite{acharya2021shadow,nguyen2022optimizing}. Up to now, Classical shadow tomography has found applications in diverse fields, including energy estimation \cite{hadfield2021adaptive}, entanglement detection \cite{neven2021symmetry,elben2020mixed},  and quantum chaos \cite{joshi2022probing}, quantum gate engineering cycle \cite{helsen2023shadow}, and quantum error mitigation \cite{seif2023shadow} to name a few.

\begin{figure}[!ht] \centering\includegraphics[width=0.4\textwidth]{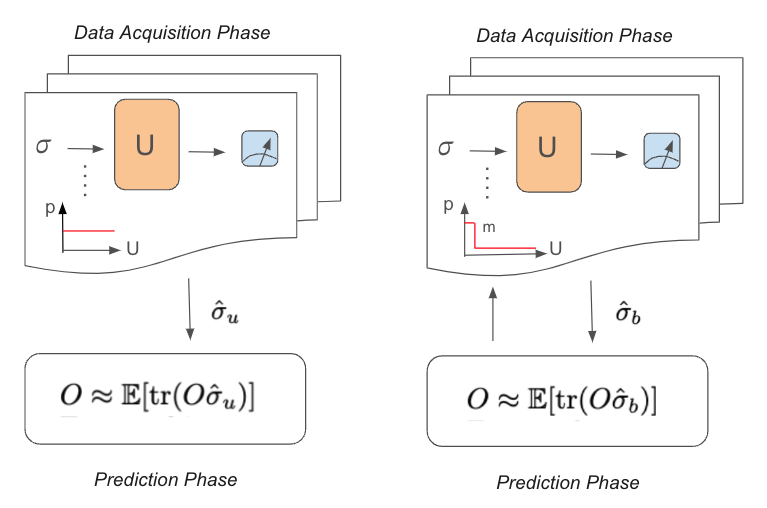}
\caption{The procedure of the classical shadow tomography.}
\label{fig:out1}
\end{figure}

Without ancilla, the minimal number of IC ensemble contains $2^n+1$ unitary operations \cite{schwinger1960unitary}. Projective measurements onto the set of $2^n+1$ mutually unbiased bases (MUBs) are recognized as the optimal approach for quantum tomography \cite{schwinger1960unitary,ivonovic1981geometrical,wootters1989optimal}.
For a vector prepared within a specific MUB, a uniform distribution will be achieved when projecting it onto any other MUBs. These MUBs measurements are regarded as maximal incompatibility and complementarity \cite{designolle2019quantifying}, finding applications in various aspects of quantum information science, including quantum tomography \cite{adamson2010improving,lima2011experimental}, uncertainty relations \cite{maassen1988generalized,ballester2007entropic,massar2008uncertainty}, quantum key distribution \cite{cerf2002security,mafu2013higher}, quantum error correction \cite{calderbank1997quantum,calderbank1998quantum,gottesman1998fault}, as well as the identification of entanglement and other forms of quantum correlations \cite{spengler2012entanglement,giovannini2013characterization,maccone2015complementarity,erker2017quantifying,kaniewski2019maximal,tavakoli2021mutually}.

In this work, we use the MUBs as the unitary ensemble for classical shadow tomography. The reconstruction channel and the variance for MUBs are computed. We find that the variance for some bounded norm observables can be exponential with the number of qubits, but it can become polynomial when the observables (or state) are approximate MUBs-average (AMA). For observables that are not AMA but obey the MUBs-sparse condition, we show that by biased sampling, one can reduce the variance to polynomial order. The procedure of the classical shadow tomography based on MUBs circuits is summarized in Figure \ref{fig:out1}. In the end, we compare the algorithmic and circuit complexity of the MUBs and Clifford measurements.

\section{MUBs Classical shadow tomography}

 Let $A$ and $B$ be two Hermitian operators with normalized eigenstates $\{|a_i\rangle\}_{i=1}^{d}$ and $\{|b_j\rangle\}_{j=1}^{d}$. These two bases are called {\it mutually unbiased} if they satisfy the property
\begin{equation} \label{eqn:mub_pp}
    \lvert \langle a_i \lvert b_j \rangle \rvert^2 = \frac{1}{d}
\end{equation}
for all $i$ and $j$.
For prime power $d$, there exist maximal ($d+1$) mutually unbiased bases (MUBs).

In $n$-qubit systems, the Hilbert space has a dimension of $d = 2^n$. Denote the $2^n+1$ MUBs as $\{\mathcal{B}_0,\cdots,\mathcal{B}_{2^n}\}$.
Consider $\mathcal{B}_0$ as the canonical basis $\{|t\rangle\}_{t=0}^{2^n-1}$.
Each additional basis is denoted as $\mathcal{B}_j=\{|e_k^j\rangle\}_{k=0}^{2^n-1}$.
Utilizing the Galois-Fourier method, all basis states $|e_k^j\rangle$ are explicitly constructed \cite{durt2010mutually}.
Recently, each unitary operation $V_j=\sum_{k=0}^{2^n-1}|e_k^j\rangle\langle k|$ has been efficiently decomposed into $O(n^2)$ elementary gates within $O(n^3)$ time, structured as $-H-S-CZ-$ \cite{yu2023efficient}.

%The assocaited Hermitian operators denoted by $\{|U_i\rangle\}_{i=0}^{d}$ have been studied in [] and recently been constructed using a polynomial number of elementary gates [Yu Wang, Dongsheng].

One of the nice properties of the MUBs is that the projective measurements onto them are informationally complete \cite{wootters1989optimal}.
By utilizing these projections, we can construct $4^n$ orthogonal operations according to the Hilbert–Schmidt inner product, $\text{tr}(A^{\dag}B)=(A,B)$. These operations can be constructed in the following manner:
\begin{itemize}
    \item The initial $2^n$ operations stem from basis $\mathcal{B}_0$, represented as $\{ |t\rangle\langle t| \}_{t=0}^{2^n-1}$.
    \item The remaining $4^n-2^n$ operations are generated from the bases $\{\mathcal{B}_j\}_{j=1}^{2^n}$. For each $j$, $2^n-1$ operations are constructed: $ \{V_j|k\rangle\langle k|V_j^{\dag}-I/2^n \}_{k=0}^{2^n-2} $.
    \label{orthogonal}
\end{itemize}

\subsection{Procedure}
We can employ MUBs circuits as a unitary ensemble for conducting classical shadow tomography \cite{huang2020predicting}.
Consider $\sigma$ as the unknown quantum state and $O$ as the observable for prediction. The general procedure encompasses two primary steps.

The initial step involves generating classical shadows of state $\sigma$ utilizing MUBs measurements.
We randomly select a $U_j$ from the $2^n+1$ MUBs circuits and rotate the unknown state, i.e., $\sigma \to U_j \sigma U_j^{\dag}$. Here, $U_j=V_j^{\dag}$. Subsequently, the qubits are measured on the computational basis. This measurement yields a 0/1 bit string of length $n$, denoted by $b_0\cdots b_{n-1}$.
Let $k=b_0+2b_1+\cdots+b_{n-1}2^{n-1}$.
We calculate the \textit{classical snapshot} of $\sigma$, defined as
\begin{equation}\label{snapshot}
    \hat\sigma = \mathcal{M}^{-1} (U_j^{\dag}| k \rangle \langle k |U_j) \;,
\end{equation}
where $\mathcal{M}^{-1}$ represents the reconstruction channel depending on the chosen unitary ensemble. Interestingly, when uniformly sampling from MUBs circuits, the reconstruction channel mirrors that of Clifford circuits and can be expressed for any operator $X$ as
\begin{equation} \label{eqn:invMmubs}
\mathcal{M}^{-1}_{\text{u}}(X) = (2^n+1)X-\text{tr}(X)I_{n} \;.
\end{equation}
The specific calculations are detailed in Appendix A.
Repeat this rotation-measurement process $N$ times. This yields a set of $N$ classical snapshots, termed a \textit{classical shadow} of $\sigma$, which will be stored in the classical memory.

The second step involves using the obtained classical shadows to predict observables $\{O_1,O_2,\ldots, O_M\}$ of the unknown quantum state $\sigma$. Their expectation values are given by
\begin{equation*}
    o_i = \text{tr}(O_i \sigma), \quad 1\leq i\leq M \;,
\end{equation*}
which can be approximated by the median of means of the expectation values
\begin{equation*}
    \hat o_i (N,K) = \text{median}\{\hat o^{(1)}_i (L,1), \hat o^{(2)}_i (L,1),\ldots, \hat o^{(K)}_i (L,1)\}
\end{equation*}
where $L = \lfloor N/K \rfloor$ and
\begin{equation*}
    \hat o^{(k)}_i (L,1) = \frac{1}{L} \sum_{j=(k-1)L+1}^{k L} \text{tr}(O_i \hat{\sigma}_j), \quad 1\leq k \leq K\;.
\end{equation*}
This approximation depends on the choices of the parameters $L$ and $K$.

\subsection{Performance}

Suppose the unknown state is $\sigma$. To assess the performance of using MUBs-based classical shadow tomography to predict an observable $O$ under $\sigma$, one should examine the variance $\lVert O_0\rVert_{\sigma}^2$ in equation (\ref{variance}), where $O_0=O-\frac{\text{tr}(O)}{2^n}$ represents the traceless part of $O$.
The sample complexity is linearly correlated with the variance.

\begin{equation} \label{variance}
    \lVert O_0 \rVert_ { \sigma}^2 = \mathbb E_{U \sim \mathcal{U}} \sum_{k=0}^{2^n-1} \langle k| U \mathcal{M}^{-1}(O_0) U^\dag | k\rangle^2 \cdot \langle k |U \sigma U^\dag | k\rangle
\end{equation}

The variance for an arbitrary unknown state is defined by the shadow norm $\lVert O_0 \rVert_\text{shadow}^2 = \max_{\sigma: \text{state}} \lVert O_0 \rVert_\sigma^2$. The variance of the shadow norm for Clifford
and Pauli measurements has been studied in \cite{huang2020predicting}.
Now, we consider the shadow variance of the MUBs measurement. Using the result of the reconstruction channel in equation \eqref{eqn:invMmubs}, it is straightforward to show that
\begin{align} \label{eq:vari1}
   \lVert O_0 \rVert_{\sigma,\text{u}}^2
& =
 (2^n+1) \sum_{j=0}^{2^n} \sum_{k=0}^{2^n-1} \text{tr}^2(O_0 P_{jk})    \cdot  \text{tr} (\sigma  P_{jk} )
\end{align}
where
$P_{jk} = U_j^{\dag}| k \rangle \langle k |U_j$.
This variance depends exponentially on $n$ because the terms $\text{tr}(\sigma P_{jk})$ and $\text{tr}(O_0P_{jk})$ can take their maximal value $\sim O(1)$ at the same $j,k$. For example, when $O=|0\rangle\langle 0|$ and $\sigma=|0\rangle\langle 0|$, the variance is $\lVert O_0 \rVert_ { |0\rangle\langle 0|,\text{u} }^2\ge (2^n+1)(1-1/2^n)^2$.
Since each $0 \le \text{tr}(\sigma P_{jk})\le 1$ for all $j,k$, we can derive that the variance $\max_{\sigma: \text{state}} \lVert O_0 \rVert_{\sigma,\text{u}}^2  \le (2^n+1) \text{tr}(O_0^2)$ for all $\sigma$ and $O$.  Thus, when the unitary ensemble in the shadow tomography is MUBs circuits, to obtain an accurate enough prediction of $\langle O \rangle$, for the worst case, one needs to perform $\sim 2^n$ many samples.

\begin{figure}[!ht] \centering
\includegraphics[width=0.45\textwidth]{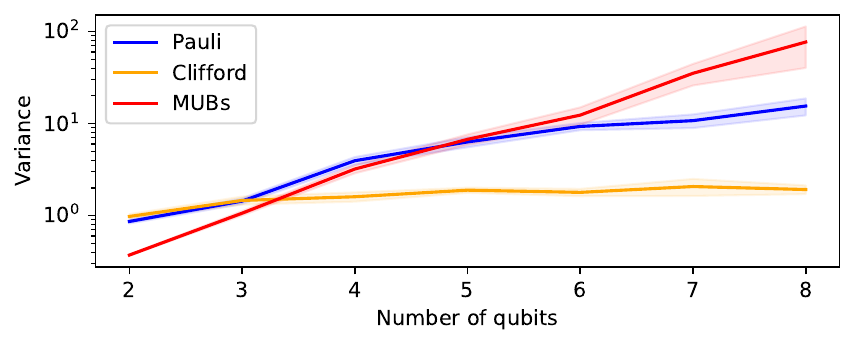}
\caption{
The variance of fidelity estimation between two GHZ states obtained from the shadows with different unitary ensembles.
For each case, we randomly generate $10$ different shadows.
The number of measurements in each shadow is $1000$.
The shaded area is the statistical variance for $10$ independent experiments. }
\label{fig:unbias}
\end{figure}

A numerical experiment is performed. Consider the observable $O=|\text{GHZ}\rangle\langle\text{GHZ}|$ on the unknown state $|\text{GHZ}\rangle\langle\text{GHZ}|$.
We predict the expectation value of $\langle O \rangle$, equivalently the fidelity between two GHZ states, for $n$ up to $8$ using shadows generated by $1000$ Pauli, Clifford, and MUBs measurements. The experiments are independently performed $10$ times. We plot the variance $\lVert O \rVert_ { \sigma}$ in Figure \ref{fig:unbias}, where the shaded area is the statistical variance between $10$ different random experiments. The results show that the variance of the prediction with shadows of Clifford measurements is independent of the number of qubits while for shadows obtained using Pauli and MUBs measurements, $\lVert O \rVert_ { \sigma}$ scales exponential with $n$, which is consistent with the analysis made above.

Although when the unitary ensemble is MUBs circuit, the variance $\lVert O_0 \rVert_{\sigma,\text{u}}$ of the prediction $\langle O \rangle$ using classical shadow tomography depends exponentially with $n$, it can be shown that this variance can decrease to polynomial with $n$ when the interested observable $O$ or the unknown state $\sigma$ has the following property.
\begin{definition}[Approximately MUBs-average]
\label{definition1}
    A state $\sigma$ (or observable $O$)  is called approximately MUBs-average (AMA) if it satisfies
    \begin{equation}
        |\text{tr}(\sigma P_{jk})-1/2^n| \le \epsilon
    \end{equation}
    for all $j,k$ and $\epsilon=O(poly(n))/2^n \ll 1$.
\end{definition}
In other words, its probability distribution under each basis of MUBs is approximately uniform. Or if we express $\sigma$ with the $4^n$ orthogonal operations defined above, the matrix elements are all less than $\epsilon+1/2^n$. Here $\epsilon$ reflects the deviation from the MUBs-average state $I/2^n$ with $\epsilon=0$. $\lVert O_0 \rVert_{I/d,\text{u}}^2=(1+1/2^n)\text{tr}(O_0^2)$. For $\epsilon = \frac{s}{2^n} \neq 0$, we can find many AMA states in the Hilbert space.

\begin{figure}[!ht] \centering
\includegraphics[width=0.4\textwidth]{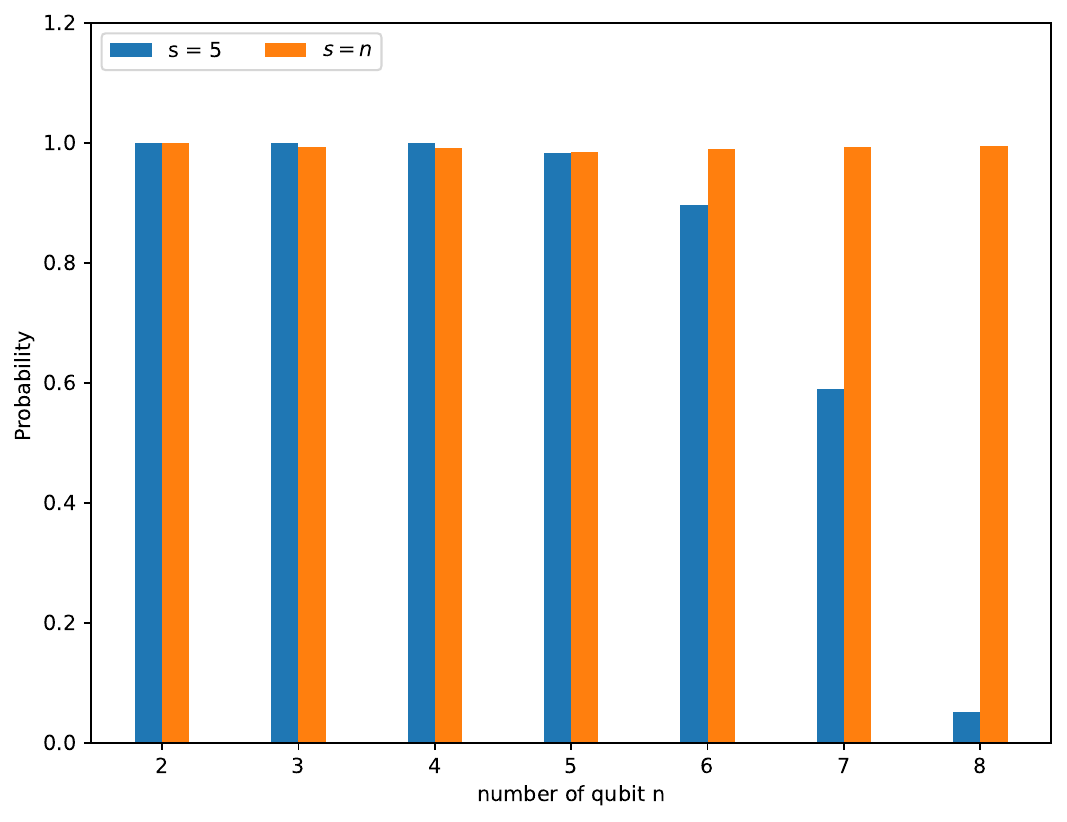}
\caption{The percentage of AMA states with $\epsilon = s/2^n$ in $1000$ randomly chosen quantum states according to Haar measure in $n=2,3, \ldots, 8$ qubit systems. }
\label{fig:bar}
\end{figure}

For example, we randomly generate $1000$ quantum states according to Haar measure in $n$-qubit system for $n=2,3,\ldots, 8$. If we set $s=5$, we find that $89.7\%$ of these random states are AMA for $n=6$, $59\%$ for $n=7$, and $5.2\%$ for $n=8$. While for $n=2,3,4,5$, almost all these random states are AMA because $s=5$ is comparable with the $2^n$. However, if we increase it to $s=n$, almost all of these random states are AMA even for $n=8$. Thus, it seems that as long as we choose an approximate $s$ such that $\epsilon$ is large enough but still $\ll 2^n$, there are a large amount of AMA states in the Hilbert space.

\begin{theorem}
\label{thm1}
If the observable $O$ is AMA, then
% We consider the observable is $O=|\phi\rangle\langle \phi|$ for fidelity estimation.
% If state $|\phi\rangle$ is approximately MUB-average,
for any unknown states $\sigma$, the upper bound of the variance is
 \begin{equation} \label{u1}
     \lVert O_0 \rVert_ { \sigma, \text{u} }^2 \le (1+\frac{1}{2^n}) poly^2 (n) \;.
 \end{equation}
On the other hand, if the unknown state $\sigma$ is AMA, then for any observable $O$ the variance is upper bounded by
\begin{equation}\label{u2}
    \lVert O_0 \rVert_ { \sigma, \text{u} }^2 \le  (1+\frac{1}{2^n})(1+poly(n)) \text{tr}(O_0^2)\;.
\end{equation}
If $\text{tr}(O_0^2)$ is constant bounded norm, then the variance is bounded by a polynomial function of $n$.
\end{theorem}
The proof of the theorem is left in Appendix B.
Thus, if the observable is AMA or the unknown state happens to be AMA, the MUBs-based shadow tomography is an effective method to predict $\langle O \rangle$.

\begin{figure}[!ht] \centering
\includegraphics[width=0.45\textwidth]{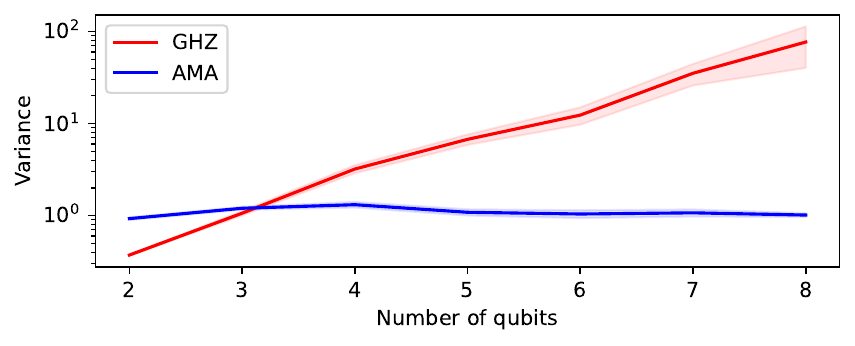}
\caption{The variance of fidelity estimation based on MUBs measurements.
The unknown state is $|\text{GHZ}\rangle\langle\text{GHZ}|$. The observable is $O_{\text{GHZ}}=|\text{GHZ}\rangle\langle\text{GHZ}|$ (red) and $O_{\text{AMA}}=|\phi\rangle \langle \phi|$ (blue) where $|\phi\rangle$ are randomly chosen AMA states with $\epsilon = 2/2^n$. The number of measurements in each shadow is $1000$.
The shaded area is the statistical variance for $10$ independent experiments.
% for two GHZ states (red) and for
% between a GHZ state and a randomly chosen AMA states with $\epsilon = n/2^n$ (Blue) and another GHZ state (red).
}
\label{fig:case2}
\end{figure}

As an example, we study the expectation value $\langle O \rangle$ on the GHZ state for both AMA observable $O_{\text{AMA}}=|\phi\rangle \langle \phi|$ and non-AMA observable $O_{\text{GHZ}}=|\text{GHZ}\rangle\langle\text{GHZ}|$ for $n=2,3,\ldots, 8$. Here, $|\phi\rangle$ are randomly chosen AMA states with $\epsilon = 2/2^n$ for each $n$.
By $1000$ MUBs measurements, we plot the variance of the prediction $\langle O \rangle$ in Figure \ref{fig:case2}. Compared with the $O_{\text{GHZ}}$ case, the variance of $O_{\text{AMA}}$ does not depends exponentially on $n$. Thus, it confirms our Theorem \ref{thm1} and one can use the MUBs-based classical shadow tomography to predict the AMA observables.

\section{Biased-MUBs classical shadow tomography}

When both the unknown state $\sigma$ and observable $O$ are not AMA, %$\text{tr}(\sigma P_{jk})\sim O(1)$ and $\text{tr}(O P_{jk})\sim O(1)$ for some $j,k$,
the variance $\lVert O_0 \rVert_ { \sigma, \text{u} }^2$   could be exponential with $n$. To obtain an accurate prediction of $\langle O \rangle$, it requires $\sim 2^n$ MUBs measurements in classical shadow tomography. While, we find that if the observable $O=|\phi\rangle\langle \phi|$ or the unknown state $\sigma$ satisfies the MUBs-sparse condition, by the biased sampling, the variance can also be $poly(n)$.

\begin{definition}[MUBs-sparse]
A state $|\phi\rangle$ is called MUBs-sparse if it has sparse expression under some basis $\mathcal{B}_j$.  Precisely, $|\phi\rangle=\sum_{k=0}^{2^n-1} a_k U_j^{\dag}|k\rangle$ for some $j$ and the nonzero elements in set $\{a_k:k=0,\cdots,2^n-1\}$ is $t=O(poly(n))$.
\end{definition}

If a state is MUBs-sparse under basis $\mathcal{B}_j$,
it is not AMA, as $\epsilon=O(1)$ under the measurement of $\mathcal{B}_j$. However, for the other $2^n$ bases, we can prove $\epsilon=t/2^n$ as follows.
%by the following Proposition \ref{MUBs sparse}.

\begin{proposition}\label{MUBs sparse}
Consider that $|\phi\rangle$ contains at most $t$ nonzero amplitudes under the expression of $\mathcal{B}_j$. For the other $2^n$ MUBs, we can prove $|\langle \phi|U_{j'}|k\rangle|^2\le t/2^n$, where $j'\ne j$.
\end{proposition}

\begin{proof} We may as well express $|\phi\rangle = \sum_{k=1}^t a_{l_k}U_{j}^{\dag}|l_k\rangle$.
With the definition of MUBs, $|\langle k'|U_{j'}  U_j^{\dag}|l_k\rangle|=\frac{1}{\sqrt{2^n}}$ for $j\ne j'$.
    \begin{align*}
    |\langle \phi|U_{j'}|k\rangle|^2 &= \frac{1} {2^n}|\sum_{k=1}^t a_{l_k} e^{i \theta_{jk}}|^2 \\
    &\le \frac{1}{2^n}(\sum_{k=1}^t |a_{l_k}|^2 )(\sum_{k=1}^t |e^{i\theta_{jk}}|^2)=\frac{t}{2^n}
\end{align*}
\end{proof}

As we will see in the following, for MUBs-sparse observables, the variance of prediction can decrease significantly if one performs the MUBs measurements in a biased way. Our motivation to consider the biased sampling is from the observations in fidelity estimation between two GHZ states. In this case, both the unknown states and observable is not AMA but they are MUBs-sparse with $t=2$ under $\mathcal{B}_0$. If we sample the MUBs circuits $\{U_i\}_{i=0}^{d}$ uniformly in shadow tomography, the variance is exponential with $n$ from the Figure \ref{fig:unbias}. In specific, $2000$ snapshots are enough to give $0.99$ fidelity for $n\le 7$, but when $n\geq 8$, ensuring a fidelity of $0.95$ requires $10000$ samplings. Upon reviewing the numerical outcomes, we find that choosing more operation $U_0$ improves the estimated performance. Theoretically, $\text{tr}(O\mathcal{M}^{-1}(P_{0k}))=(2^n-1)/2$  when $U_0$ is chosen. If not, the expectation values could be $1/2^{n+1}-1/2$ or $1/2^n$. However, as $n$ exceeds 8, the likelihood of uniformly selecting $U_0$ drastically decreases. Appropriately increasing the number of samples of $U_0$ will result in a faster approximation of the exact expected value $1$.

Biased shadow tomography follows a similar process to the usual case, with the difference lying in the data acquisition phase—employing random MUBs measurements based on a biased distribution. For instance, if the observable $O=|\phi\rangle\langle\phi|$ is  MUBs-sparse under $\mathcal{B}_j$, adjusting probabilities can prioritize sampling from $U_j$ while reducing others' likelihood, as outlined below.

\begin{align} \label{eqn:biasedP}
     p_{U_j} = \frac{1}{2^n+1} \to \frac{1+m}{2^n+1+m}, \qquad  &    U_j \nonumber \\
     p_{U_k} = \frac{1}{2^n+1} \to \frac{1}{2^n+1+m},   \qquad &  \text{otherwise}
\end{align}
where $m$ is a real number. Note that when we use the biased shadow to predict $\langle O \rangle$, the variance for uniform sampling in equation (\ref{eq:vari1}) does not apply anymore. In fact, the correct variance depends on the parameter $m$. As we will see later, $m$ is not a free parameter in our scheme but will be fixed to be the one that minimizes this variance.

When we randomly sample the MUBs circuits according to equation \eqref{eqn:biasedP}, for any operator $X$, the reconstruction channel becomes
\begin{equation}
\mathcal{M}_b^{-1}(X)=(2^n+1+m)[X-\frac{m}{1+m} \text{tr}(X P_{jk})P_{jk} ]- \frac{\text{tr}(X)I}{1+m}.
\end{equation}
When $m=0$, it becomes the reconstruction channel for the uniform sampling case in equation \eqref{eqn:invMmubs} and $\mathcal{M}_b^{-1}(X)=\mathcal{M}_u^{-1}(X)$.
The details of this derivation can be found in Appendix \ref{sec:app1}.

\begin{theorem} \label{thm2}

Given an observable $O=|\phi\rangle\langle \phi|$ and unknown quantum state $\sigma$, if $|\phi\rangle$ is MUBs-sparse, one can efficiently predict   $\langle O \rangle$ using biased MUBs sampling. The upper bound of variance is given by \begin{equation} \label{eqn:biased-vari1}
    \lVert O_0 \rVert_{\sigma,b}^2 \le   t^2, \; \text{~when~} \; m=\frac{2^n}{t-1}-1.
\end{equation}
and the optimal probabilities to choose $U_j$ and the other unitary circuits in the MUBs-set are
\begin{equation} \label{sampleprobability}
    p_{U_j} = \frac{1}{t};\qquad p_{U_{k}} = \frac{1}{2^n}\left(1 - \frac{1}{t}\right), \quad k\ne j
\end{equation}
On the other hand, given an observable $O$ and unknown quantum state $\sigma=|\phi\rangle\langle \phi|$, if $|\phi\rangle$ is MUBs-sparse, one can   predict   $\langle O \rangle$ using biased MUBs sampling. The upper bound of variance is given by

\begin{equation} \label{eqn:biased-vari2}
    \lVert O_0 \rVert_{\sigma,b}^2 \le   (\sqrt{t}+1)^2 \text{tr}(O_0^2), \; \text{when } \; m=\frac{2^n}{\sqrt{t}}-1.
\end{equation}
and the optimal probabilities to choose $U_j$ and the other unitary circuits in the MUBs-set are
\begin{equation} \label{sampleprobability}
    p_{U_j} = \frac{1}{1+\sqrt{t}};\qquad p_{U_{k}} = \frac{1}{2^n}\left( \frac{1}{1+\sqrt{t}}\right), \quad k\ne j
\end{equation}
\end{theorem}

The proof is left in the Appendix B. For both cases, the variance for predicting the MUBs-sparse observable is upper bounded by a function independent of $n$. Thus, classical shadow tomography based on the MUBs circuits works not only for AMA observables but also for observables that are MUBs-sparse by biased sampling.

\begin{figure}[!ht] \centering
\includegraphics[width=0.45\textwidth]{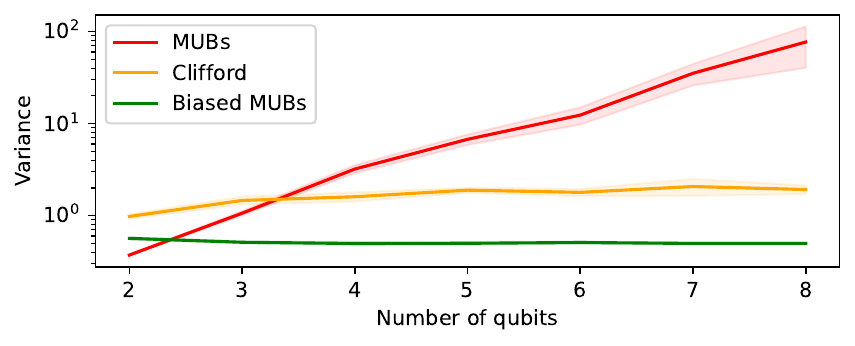}
\caption{The variance of fidelity estimation between two GHZ states using MUBs shadows (red), Clifford shadows (orange), and Biased MUBs shadows (green).
The shaded region is the statistical variance among $10$ randomly generated shadows.
% For each case, we randomly generate $10$ different shadows.
The number of measurements in each shadow is $1000$.
}
\label{fig:bias}
\end{figure}

Let's consider the fidelity estimation between two GHZ states again. In Figure \ref{fig:bias}, we compare the variance of the predictions of $O_{\text{GHZ}}$ using the shadows obtained by Clifford, MUBs and biased MUBs measurements. As discussed before, the observable is not AMA. Uniform sampling in the shadow tomography leads to the exponential dependence of $n$ in the variance. While since $O_{\text{GHZ}}$ is MUBs-sparse with $t=2$ under $U_0=I$, according to Theorem \ref{thm2}, we perform the MUBs measurements in a biased ways with the probability $0.5$ on $U_0$ and $1/2^{n+1}$ on the others. By doing so, the variance decreases significantly that is even lower than the Clifford case. In this simple example, we have shown that by biased sampling, the classical shadow tomography based on MUBs circuits works for the MUBs-sparse observables.

\begin{remark}
    Note that for observables that can be treated as the density matrix of mixed states i.e. $O=\sum_{k=1}^s p_k |\phi_k\rangle\langle \phi_k|$ with $p_k \ge 0 $ and $\sum_{k=1}^s p_k =1$, if all $\{|\phi_k\rangle\}$ are MUBs-sparse under the same basis $\mathcal{B}_j$, then the theorem \ref{thm2} still holds.
\end{remark}

It is easy to check that $\text{tr}(OP_{j'k})\le \frac{t}{2^n}$ for each $k$ when $j'\ne j$.

\section{Comparison with random Clifford measurements}
\begin{figure}[!ht] \centering\includegraphics[width=0.25\textwidth]{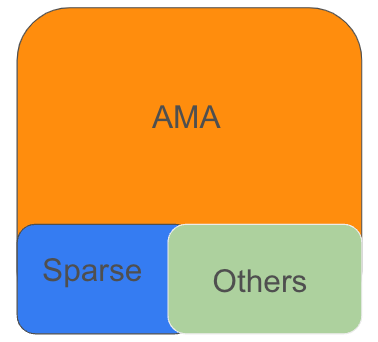}
\caption{States classification. We divide all $n$-qubit states into three parts, AMA states (observables), states (observables) with a `sparse' representation under some basis $\mathcal{B}_j$, and others. It's intriguing to explore if other states share similar properties with the sparse states and to assess the percentage of these failed states within the complete set.}
\label{fig:ABC}
\end{figure}

Based on the performance of the classical shadow tomography with MUBs ensemble, we classify the quantum states (observables) into three classes as depicted in Figure (\ref{fig:ABC}). With uniform MUBs circuit sampling, we observe that the worst-case variance is bounded by $(2^n+1)\text{tr}(O_0)^2$, while the average variance is $(1+1/2^n)\text{tr}(O_0)^2$. Notably, for AMA states (observables), the variance remains polynomial. Additionally, with biased MUBs circuit sampling, the variance for sparse states (observables) also stays polynomial. By contrast, when employing uniform sampling of Clifford circuits, the variance for all states
(observables) is limited to $3\text{tr}(O_0^2)$.

In addition to the performance, experiment implementation and post-processing are two main procedures in classical shadow tomography.
We compare them for Clifford ensembles and MUBs ensembles.

\paragraph{Experiment implementation}

In the data acquisition phase, one needs to randomly sample the unitary operations in the ensemble many times. The MUBs ensemble has much fewer elements compared with the Clifford one, thus making the sampling process more direct. Also, the random MUBs circuit structure is simpler to implement in the experiments than the random Clifford circuit. Surprisingly, we only need to consider $n$ special MUBs circuit to implement all.

\begin{itemize}
    \item Sampling process

\hspace{0.3cm} As for the $n$-qubit Clifford ensemble, the number of elements
is $2^{n^2+2n}\prod_{j=1}^{n} (4^j-1)$. The quantity is too large to sample directly from the first to the last. Nevertheless, each Clifford circuit $U$ is fully characterized by its action on $2^n$ Pauli operators \cite{gottesman1997stabilizer}:  $U X_j U^\dagger = (-1)^{r_j} \prod_{i=1}^{n} X_i^{\alpha_{ji}} Z_i^{\beta_{ji}}$, and $U Z_j U^\dagger = (-1)^{s_j} \prod_{i=1}^{n} X_i^{\gamma_{ji}} Z_i^{\delta_{ji}}$. The parameters that define $U$ are $(\alpha, \beta, \gamma, \delta, r, s)$, where $\alpha, \beta, \gamma, \delta$ are $n \times n$ matrices of bits, and $r, s$ are $n$-bit vectors.
Given these parameters, different methods can decompose $U$ into elementary circuits \cite{gottesman1997stabilizer,aaronson2004improved,koenig2014efficiently,maslov2018shorter,van2021simple,bravyi2021hadamard}. The time complexity is $O(n^3)$ or $O(n^2)$, while the number of elementary gates is $O(n^2/\log n)$ or $O(n^2)$.

\hspace{0.3cm} As for the $n$-qubit MUBs ensemble, there are $2^n+1$ circuits.
This allows for a direct uniform (biased) sampling due to their significantly lower count compared to Clifford circuits.
In paper \cite{yu2023efficient}, $2^n+1$ circuits $\{I,U(0),U(1),\cdots,U(2^n-1)\}$ produce $2^n+1$ MUBs by acting on computational basis.
The sampled MUBs circuits are  $\{I,U^{\dag}(0),\cdots,U^{\dag}(2^n-1)\}$.
Each nontrivial circuit $U(j)$ is obtained within $O(n^3)$ time.
Notably, every circuit comprises at most $(n^2+7n)/2$ elementary gates. On average, the counts of gates includes $3n/2$ $S$ gates, $(n^2-n)/4$ $CZ$ gates, and $(n-u)/2$ $CZ$ gates with a distance $u$.

    \item Circuits structure

\hspace{0.3cm}  For the Clifford ensemble, the circuit is structured with   11-stage decomposition: $-H-CX-S-CX-S-CX-H-S-CX-S-CX-$ \cite{aaronson2004improved}, or  7-stage   decomposition: $-CX-CZ-S-H-S-CZ-CX-$ \cite{maslov2018shorter,bravyi2021hadamard}.

\hspace{0.3cm}  For the MUBs ensemble, the circuit is structured by $-H-S-CZ-$.
The entanglement $CZ$ component consists of $2n-3$ fixed modules.
Furthermore, the structures among different MUBs circuits exhibit a strong correlation. Linear relations exist in these $2^n+1$ MUBs circuits. Each $U(j)$ can be derived from the following $n$ circuits: $\{U(2^0),U(2^1),\cdots,U(2^{n-1})\}$.

\end{itemize}

\paragraph{Post-processing}

In the prediction phase, we require two sets of data: the unitary operations used in the rotation, $U_j$, and the measurement outcomes, $|k\rangle$. With them, one can estimate the expected value of observable $O$ by
\begin{equation}
   o \approx \text{tr}[\mathcal{M}^{-1}(U_j^{\dag}|k\rangle\langle k|U_j) O] \;.
\end{equation}
However, as $n$ grows, the evaluation of the above expression becomes exponentially slow.
This challenge is evident for Clifford circuits when the observable $O$ lacks efficient representations, such as an efficient stabilizer decomposition. For MUBs circuits, a similar issue may arise when $O$ lacks a sparse representation under some basis $\mathcal{B}_j$.

In the case of Clifford ensembles, the number of all stabilizer states ${U_j^{\dag}|k\rangle}$ is $O(2^{n^2/2})$, while the number of MUBs states is $2^n(2^n+1)$. Consequently, the number of possible classical snapshots in equation (\ref{snapshot}) decreases for MUBs ensembles. Although the set of all MUBs states is a subset of all stabilizer states, the Clifford ensembles prove effective for a broader range of observables, considering the computational complexity involved in obtaining the estimation $o$.

The circuit decomposition can be realized in all physical platforms.
In a quantum optical experiment, one could perform the random Clifford measurements by uniformly projecting a stabilizer state $U^{\dag}_j|k\rangle$. To compute all these $2^n$ coefficients, the time complexity is $O(n\cdot 2^{3n})$ when we use the exression of stabilizers generators $\{g_i\}_{i=1}^n$, where  $U^{\dag}_j|k\rangle\langle k|U^{\dag}_j=\frac{1}{2^n}\prod_{i=1}^{n}(I+g_i)$. A new approach reduces the complexity to $O(2^n n^3)$ \cite{struchalin2021experimental}. If we randomly project a MUBs state $U(j)|k\rangle$, the time complexity is also  $O(2^n n^3)$ and one can directly calculate the coefficients \cite{durt2010mutually,yu2023efficient}.
\begin{equation*}
   U(j)|k\rangle= |f_k^j\rangle=\frac{1}{\sqrt{2^n}}\sum_{l=0}^{2^n -1}|l\rangle(-1)^{k\cdot l^T}\cdot \alpha_l^j
\end{equation*}
where $\alpha_l^j=\prod_{s,t=0}^{n-1}\overline{(\sqrt{-1})^{j\odot (l_s\cdot 2^s)\odot(l_t\cdot 2^t)}}$ and $k$ ranges from $0$ to $2^n-1$.

\section{Conclusion}

In this paper, we explore classical shadow tomography by uniform and biased sampling MUBs circuits. The $2^n+1$ MUBs is the minimal and optimal set for full $n$-qubit state tomography, which also constitutes a subset of all Clifford circuits. MUBs circuit is structured with 3-stage decomposition  $-H-S-CZ-$ \cite{yu2023efficient}, a part of Clifford circuit \cite{bravyi2021hadamard}.  There are `linear' relations between these MUBs circuits and the average number of different gates can be counted.

The reconstruction channel and variance are calculated for random uniform and biased sampling of MUBs circuits. For the most general observable, the variance is bounded by $O(2^n)$, but if one considers a special subset defined as AMA observables, we show that the upper bound of the variance becomes $poly(n)$ which is comparable with the Clifford case. Furthermore, we find that by biased sampling of MUBs circuits, we can effectively decrease the variance to $poly(n)$ when the observable (or the unknown state) is MUBs-sparse. All these results are demonstrated by the numerical experiments.

There are many future directions that one can pursue with the classical shadow tomography with MUBs circuits. First, we would like to find an efficient scheme to predict the general observables that are not AMA and MUBs-spares. Second, it's interesting to consider more types of observables like two-point correlation functions and OTOCs. Third, one can also study how to use the MUBs circuits to predict nonlinear or polynomial observables such as entanglement entropy.

\textbf{Acknowledgements---}
The work of Yu Wang received support from the National Natural Science Foundation of China through Grants No. 62001260 and No. 42330707, as well as from the Beijing Natural Science Foundation under Grant No. Z220002. The work of Wei Cui is supported by the fellowship of China Postdoctoral Science Foundation NO.2022M720507 and in part by the Beijing Postdoctoral Research Foundation.

\onecolumngrid
\appendix

\section{Computation details of the reconstruction channel} \label{sec:app1}

In the classical shadow tomography, after a MUBs measurements $\{U_j^{\dag}|k\rangle \langle k|U_j\}_{k=0}^{2^n-1}$, the unknown density matrix $\rho$ can be viewed as collapsing to $P_{jk} =U_j^{\dag}|k\rangle \langle k|U_j$ with $|k\rangle$ a state in the computational basis. To obtain the classical shadows $\hat \rho$, one needs to know the reconstruction channel $\mathcal{M}^{-1}$. We calculate the reconstruction channels of the MUBs measurements for both uniform and biased samplings. The computation details are given below.

\paragraph{Uniform sampling}

The channel of the MUBs measurements is defined in the following way.
\begin{equation}
\mathcal{M}_{\text{u}}(\rho)
% &= \frac{1}{{2^n + 1}} \sum_{t=1}^{2^n(2^n+1)} \mbox{tr}(\rho |\phi_t\rangle\langle \phi_t|) \cdot |\phi_t\rangle\langle \phi_t|  \mbox{;~~randomly projected to $(2^n+1)2^n$ MUBs states.}\\
 =\frac{1}{{2^n + 1}} \sum_{j=0}^{2^n}\sum_{k=0}^{2^n-1} \mbox{tr}(\rho P_{jk}) \cdot P_{jk}
\label{rhosum}
\end{equation}
As the MUBs circuits are informationally complete, each $\rho$ can be expressed with the form of
\begin{equation*}
\rho=\sum_{j=0}^{2^n}\sum_{k=0}^{2^n-1}x_{jk} P_{jk}.
\end{equation*}
Note here the coefficients $\{x_{jk}\}$ may not be unique. It is straightforward to show that
\begin{align}
\mbox{tr}(\rho P_{jk}) &= \mbox{tr}[(\sum_{a=0}^{2^n}\sum_{b=0}^{2^n-1}x_{ab}P_{ab})\cdot P_{jk}]  \\
&= (\sum_{a=j,b=k}+\sum_{a=j,b\ne k}+\sum_{a\ne j}) \mbox{tr}[( x_{ab} P_{ab}\cdot P_{jk}] \\
&=x_{jk}+\underbrace{0  + \ldots + 0}_{2^n-1} +\frac{1}{2^n}\sum_{a\ne j}\sum_{b=0}^{2^n-1}x_{ab} \label{mub}\\
&=x_{jk} +\frac{1}{2^n}\sum_{a=0 }^{2^n}\sum_{b=0}^{2^n-1}x_{ab}-\frac{1}{2^n}\sum_{b=0}^{2^n-1}x_{jb}\\
&=x_{jk}+\frac{1}{2^n}\mbox{tr}(\rho)-\frac{1}{2^n}\sum_{b=0}^{2^n-1}x_{jb} \label{p}
\end{align}
Here in equation (\ref{mub}) we use the property in equation  (\ref{eqn:mub_pp}), i.e. the square of the inner product of each two eigenstates in different MUBs is $1/2^n$.
\begin{align}
\mathcal{M}_{\text{u}}(\rho) &= \frac{1}{{2^n + 1}} \sum_{j=0}^{2^n}\sum_{k=0}^{2^n-1} (x_{jk} +\frac{\text{tr}(\rho)}{2^n}-\frac{1}{2^n}\sum_{b=0}^{2^n-1}x_{jb})\cdot P_{jk}  \\
&=\frac{1}{{2^n + 1}} (\rho+\frac{\text{tr}(\rho)(2^n+1)I}{2^n}-\sum_{j=0}^{2^n}\sum_{b=0}^{2^n-1}\frac{x_{jb}}{2^n}I)\\
&=\frac{1}{{2^n + 1}} (\rho+\frac{\text{tr}(\rho)(2^n+1)I}{2^n}-\frac{\mbox{tr}(\rho)}{2^n}I)\\
&=\frac{1}{{2^n + 1}} (\rho+\text{tr}(\rho)I)
\end{align}
% Here for each $j$, $\sum_{k=1}^{2^n}V_j|k\rangle\langle k|V_j^{\dag}=I$;  $\mbox{tr}(\rho)=1$.
Then the inverse channel is given by
\begin{equation}
\mathcal{M}_{\text{u}}^{-1}(\rho)=(2^n+1)\rho-\text{tr}(\rho) I.
\end{equation}
It is the same as the reconstruction channel for the Clifford measurements \cite{huang2020predicting}.

\paragraph{Biased sampling}
Without loss of generality, we may as well let the biased sampling basis be $\mathcal{B}_0$. The sampling probability is $\frac{1+m}{2^n+1+m}$ for $U_0=I$, and the sampling probability for other $U_j$ is $\frac{1}{2^n+1+m}$ where $j\ne 0$.  The resulting reconstruction channel for this adjusted sampling process will be
  \begin{align}
\mathcal{M}_{\text{b}}(\rho)
&=\frac{1}{{2^n + 1+m}}  [\sum_{j=0}^{2^n}\sum_{k=0}^{2^n-1} \mbox{tr}(\rho P_{jk}) \cdot P_{jk} + m \sum_{k=0}^{2^n-1} \mbox{tr}(\rho P_{0k}) \cdot P_{0k} ] \\
&= \frac{1}{{2^n + 1+m}}  [\rho+\text{tr}(\rho)I +m \sum_{k=0}^{2^n-1} \mbox{tr}(\rho P_{0k}) \cdot P_{0k} ]
\end{align}

Then the updated inverse channel will be

\begin{equation}
\mathcal{M}_{\text{b}}^{-1}(\rho)=(2^n+1+m)[\rho-\frac{m}{1+m}\sum_k\text{tr}(\rho P_{0k})P_{0k}]- \frac{\text{tr}(\rho)I}{1+m}.
\end{equation}

\section{Computational details on the variance}

The performance of the classical shadow tomography is linear dependent with the variance defined below
\begin{align}
\lVert O -\frac{\text{tr}(O)}{2^n} \rVert_\text{ shadow} & = \max_{\sigma: \text{ state }}
 \left(\mathbb E_{U \sim \mathcal{U}} \sum_{b \in \{0, 1\}^n}   \langle b |U  \sigma U^\dag  | b\rangle \cdot \langle   b| U   \mathcal{M} ^{-1}(O -\frac{\text{tr}(O)}{2^n}) U^\dag | b\rangle ^2 \right)^{1/2}
\end{align}
where $U$ is a randomly chosen circuit in the unitary ensemble $\mathcal{U}$. From the definition, the variance depends on the observable $O$ and choice of the $\mathcal{U}$, but it also depends implicitly on how one samples the circuits. In this section, we calculate this variance when $\mathcal{U}$ is MUBs circuits for both uniform and biased sampling.

\paragraph{Uniform sampling}
 For the traceless part $O_0=O-\frac{\text{tr}(O)}{2^n}$, we know $\text{tr}(O_0)=0$.
The reconstruction channel   is
$$\mathcal{M}_{\text{u}}^{-1}(O_0)=(2^n+1)O_0-\text{tr}(O_0)I=(2^n+1)O_0$$
  One obtains that
\begin{align}
   \lVert O_0 \rVert_\text{ shadow}^2 &= \max_{\sigma: \text{ state }}  \lVert O_0 \rVert_{\sigma, \text{u} }^2
   \\& = \max_{\sigma: \text{ state }}
  \frac{1}{2^n+1}\sum_{j=0}^{2^n} \sum_{k=0}^{2^n-1} \text{tr}(\sigma P_{jk})\cdot \text{tr} ^2
  ((2^n+1)O_0 P_{jk})  \\
& =  \max_{\sigma: \text{ state }}
 (2^n+1) \sum_{j=0}^{2^n} \sum_{k=0}^{2^n-1} \text{tr}^2(O_0P_{jk})    \cdot  \text{tr} (\sigma  P_{jk} )
\end{align}
%

% \begin{proposition}
% $0\le \text{tr}(\sigma P_{jk}) \le 1$, $\sum_{j,k}\text{tr}(\sigma P_{jk})=2^n+1$.
% \end{proposition}

\begin{proposition}
% $\sum_{j,k}\text{tr}(O_0 P_{jk})=0$,

$\sum_{j,k}\text{tr}^2(O_0P_{jk})=\text{tr}(O_0^2)$.
\end{proposition}

\begin{proof}
    %Now we analysis the value of $\text{tr}(O_0P_{jk})$.
By equation (\ref{rhosum}), we have
\begin{align}
    &\sum_{j=0}^{2^n} \sum_{k=0}^{2^n-1} \text{tr}(O_0 P_{jk})    \cdot P_{jk}=(2^n+1)\mathcal{M}({O_0})
    =O_0-\text{tr}(O_0)I=O_0
\end{align}
Thus $\sum_{j,k}\text{tr}(O_0 P_{jk})=\text{tr}(O_0)=0$.

Define $O_{0,j}=\sum_{k=0}^{2^n-1}\text{tr}(O_0P_{jk}) P_{jk}$, thus $O_0=\sum_j O_{0,j}$.
It is easy to prove $\text{tr}(O^2_{0,j})=\sum_{k=0}^{2^n-1}\text{tr}^2(O_0P_{jk}) $.
If $j\ne j'$, we have
\begin{align}
   \text{tr}(O_{0,j}O_{0,j'}) &=\frac{1}{2^n}\sum_{k,k'}\text{tr}(O_0P_{jk})\text{tr}(O_0 P_{j'k'})\\
   &=\frac{1}{2^n}\text{tr}(O_0\sum_kP_{jk})\text{tr}(O_0\sum_{k'} P_{j'k'})\\
   &=0
\end{align}
Here we use $\text{tr}(O_0I)=0$.
Thus $\text{tr}(O_0^2)=\sum_{j}\text{tr}(O^2_{0,j})=\sum_{j}\sum_{k}\text{tr}^2(O_0P_{jk})$.
\end{proof}

Now, we give the proof of equation (\ref{u1}).
\begin{proof}
As $\text{tr}(\sigma P_{jk})\le \epsilon+\frac{1}{2^n}$ for all $j,k$, we have  $\lVert O_0 \rVert_ { \sigma,\text{u} }^2 \le (2^n+1)(\epsilon+\frac{1}{2^n})\sum_{j,k}\text{tr}^2(O_0P_{jk})  =(2^n+1)(\epsilon+\frac{1}{2^n}) \text{tr}(O_0^2)$.
Thus  $\lVert O_0 \rVert_ { \sigma,\text{u} }^2\le (1+\frac{1}{2^n})(1+poly(n))\text{tr}^2(O_0)$.

  When $\epsilon=O(1/2^n)$ and $\text{tr}(O_0^2)$ is bounded,   the variance for these states $\sigma$ is at  a constant level.
 When  $\sigma=I/2^n$ we have $\epsilon=0$ and $\lVert O_0 \rVert_ { \sigma }^2 \le \frac{2^n+1}{2^n} \text{tr}(O_0^2)$.
\end{proof}

Here, we give the proof of equation (\ref{u2}).

\begin{proof}
    As $\text{tr}(O_0 P_{jk})=\text{tr}(O P_{jk})-1/2^n$, we have $\text{tr}^2(O_0 P_{jk})\le \epsilon^2$.
    Then $\lVert O_0 \rVert_ { \sigma,\text{u} }^2
     \le  (2^n+1) \epsilon ^2 \cdot \sum_{j,k}\text{tr}(\sigma P_{jk})= (2^n+1)^2 \epsilon ^2$. Thus
      $\lVert O_0 \rVert_ { \sigma, \text{u} }^2 \le (1+\frac{1}{2^n}) poly^2 (n)$.
\end{proof}

\paragraph{Biased sampling}

Here we give the proof of Theorem 2.
 \begin{proof}
     The inverse reconstruction channel for biased sampling is as follows,
\begin{equation}
    \mathcal{M}_{{\text{b}}}^{-1}(O_0)=(2^n+1+m)[O_0-\frac{m}{1+m}   \sum_k\text{tr}(O_0P_{0k})P_{0k}]
\end{equation}

Denote $O_{0,m}=O_0-\frac{m}{1+m}   \sum_k\text{tr}(O_0P_{0k})P_{0k}$.

The variance for state $\sigma$ changes to  the following:
\begin{align*}
  \lVert O_0 \rVert_{\sigma,\text{b}}^2
    &   = (2^n+1+m)^2
  \sum_{j=1}^{2^n} \sum_{k=0}^{2^n-1} \text{tr} ^2(O_{0,m} P_{jk})  \cdot  \frac{\text{tr} (\sigma  P_{jk} )}{2^n+1+m }
  \\
  &+(2^n+1+m)^2\sum_{k=0}^{2^n-1} \frac{\text{tr} ^2(O_{0,m}  P_{0k} )}{(1+m)^2}  \cdot  \frac{(m+1) \text{tr} (\sigma  P_{0k} )}{2^n+1+m}
\end{align*}

As we have the following relations.
When $j= 0$, $ \text{tr}(O_{0,m}P_{0k})  =\frac{1}{1+m} \text{tr}(O_0P_{0k}) $.

When $j\ne 0$,
\begin{align}
    \text{tr} (O_{0,m}  P_{jk} )&=\text{tr} (O_{0}  P_{jk} )- \frac{1}{2^n}\frac{m}{1+m}\sum_{k'}\text{tr}(O_0P_{0k'})\\
    &=\text{tr} (O_{0}  P_{jk} )
\end{align}
As $\text{tr}(O_0 I)=0$.

We can rewrite the variance as
\begin{align}
  \lVert O_0 \rVert_{\sigma,\text{b}}^2
    &   = (2^n+1+m)
  \sum_{j=1}^{2^n} \sum_{k=0}^{2^n-1} \text{tr} ^2(O_{0} P_{jk})  \cdot  \text{tr} (\sigma  P_{jk} )
  \\
  &+\frac{1}{ 1+m }(2^n+1+m)\sum_{k=0}^{2^n-1} \text{tr} ^2(O_{0}  P_{0k})  \cdot  \text{tr} (\sigma  P_{0k} )
\end{align}

%Then for all $\sigma$, we have the following relation
Now we give the proof of equation (\ref{eqn:biased-vari1}).

Given the observable $O=|\phi\rangle\langle\phi|$ and any unknown state $\sigma$, we know $\text{tr}(O_0P_{0k})\le \max_{ |k\rangle} ( \text{tr}(OP_{0k})-\frac{1}{2^n})\le 1$ and $\text{tr}(O_0P_{jk})^2\le  \frac{(t-1)^2}{2^n}$ for $j\ne 0$.
\begin{equation} \label{eqn:biased-vari}
    \lVert O_0 \rVert_{\sigma,\text{b}}^2 \le  (2^n+1+m)[\frac{(t-1)^2}{2^n\cdot 2^n}\cdot 2^n +\frac{1}{ 1+m  } \cdot  \max_{ |k\rangle} ( \text{tr}(OP_{0k})-\frac{1}{2^n})^2] \le   (2^n+1+m)[\frac{(t-1)^2}{2^n} +\frac{1}{ 1+m } ]
\end{equation}
%Here we use the property of MUBs and Cauchy-Schwarz inequality.
\end{proof}

Now we give the proof of equation (\ref{eqn:biased-vari2}).
Given the observable $O$ and unknown state $\sigma=|\phi\rangle\langle \phi|$, we know $\text{tr}(\sigma P_{0k})\le 1$ and $\text{tr}(\sigma P_{jk})\le \frac{t}{2^n}$ for $j\ne 0$.

\begin{equation}
    \lVert O_0 \rVert_{\sigma,\text{b}}^2 \le  (2^n+1+m)[\frac{t}{2^n}\cdot \sum_{j=1}^{2^n}\text{tr}(O_{0,j}^2) +\frac{1}{ 1+m  } \cdot \text{tr}(O_{0,0}^2) ] \le   (2^n+1+m)[\frac{t}{2^n} +\frac{1}{ 1+m } ]\text{tr}(O_{0 }^2)
\end{equation}

\bibliographystyle{unsrt}
\bibliography{myreferences}

\end{document}